\documentclass{iopart}
\usepackage{bm} 
\usepackage{graphicx}

\begin{document}

\title{On the theory of cavities with point-like perturbations. Part II: Rectangular cavities}
\author{T~Tudorovskiy$^{1,2}$, U~Kuhl$^{1,3}$, H-J~St\"ockmann$^1$}
\address{$^1$ Fachbereich Physik der Philipps-Universit\"at Marburg, D-35032 Marburg, Germany}
\address{$^2$ Institute for Molecules and Materials, Radboud University of Nijmegen, Heyendaalseweg 135, 6525AJ Nijmegen, The Netherlands}
\address{$^3$ Laboratoire de Physique de la Mati\`{e}re Condens\'{e}e, CNRS UMR 6622, Universit\'{e} de Nice Sophia-Antipolis, 06108 Nice, France}
\ead{t.tudorovskiy@science.ru.nl}

\date{\today}

\begin{abstract}
We consider an application of a general theory for cavities with point-like perturbations for a rectangular shape. Hereby we concentrate on experimental wave patterns obtained for nearly degenerate states. The nodal lines in these patterns may be broken, which is an effect coming only from the experimental determination of the patterns. These findings are explained within a framework of the developed theory.
\end{abstract}

\submitto{\JPA}

\maketitle

\section{Introduction}
\label{sec:intro}

Rectangular billiard perturbed by a zero-range perturbation (\v{S}eba billiard \cite{seb90}) has been investigated theoretically \cite{seb90,dem88,alb88,seb91,haa91b,che96b,bog01a,bog02a,rah02,tud08,tud1} and experimentally in microwvave experiments with cavities \cite{haa91b,kuh00b,tud08,tud1}. Experimental wavefunctions are extracted from either the resonance shift induced by a movable perturber  \cite{sri91,wu98,dem00a,lau07} or obtained by a Lorentzian fit of the reflection amplitude measured by a movable antenna \cite{ste92,ste95,kuh07b}. However these methods work only for well isolated resonances. In this paper we discuss the experimental wavefunctions in the case of nearly degenerate states. The starting point is the general expression for the scattering matrix ($S$-matrix) for such systems, that we derived in the previous paper \cite{tud08}, called Part I. It considered the main properties of cavities with point-like perturbations, following the ideas of \cite{exn97}. In the derived expression the whole information on the antenna construction was reduced to four effective constants. We have shown that a lot of experimental results can be treated within the framework of the developed approach. We present the general expression for the $S$-matrix obtained in Part I in Section~\ref{sec:sketch} of this part.

The practical application of the derived expression for $S$-matrix requires to compute the Green function of the unperturbed cavity as well as the so-called renormalized Green function \cite{dem88}. In the general case the exact computation of these functions is by no means a trivial task. In \cite{tud1} we have shown that for a rectangular cavity the required computation can be performed by an application of Ewald's method \cite{ewa21,duf01}. This observation allowed us to revise the level-spacings statistics of \v{S}eba billiard \cite{seb90}. Ewald's method gives the representation of the (renormalized) Green function in terms of exponentially convergent series. This representation fits very well for numerical studies.

The main purpose of the present Part II is a further application of the theory developed in Part I to the theoretical description of 
experimental data obtained in rectangular cavities with particular emphasis to wave functions. 
The experimental set-up and results are explained in Section~\ref{Sec:Exp}. As it was explained above, 
the main advantage of rectangular cavities is the possibility of the exact computation of the $S$-matrix even for considerably high frequencies. 

In Section~\ref{sec:WaveProbing} of this Part we treat our experimental findings \cite{kuh00b} (see also figure~\ref{fig:exp}) obtained in the experiment with two antennas in the case of nearly degenerate states. We show, that while the nodal lines of eigenfunctions of a cavity are either closed or terminate at the boundaries, the nodal lines of the experimental patterns measured in this case may break inside the cavity and reduce to points in the case of degenerate states. This unavoidable experimental imperfection was not pointed out earlier. The results of Section~\ref{sec:WaveProbing} are universal and remain valid for cavities with arbitrary shapes.

\section{Experimental set-up and results}
\label{Sec:Exp}

\begin{figure}
\begin{center}
\begin{minipage}{50mm}
 \parbox[t]{5mm}{(a)}\parbox[t]{50mm}{\ \\[-3.5ex]\includegraphics[width=50mm]{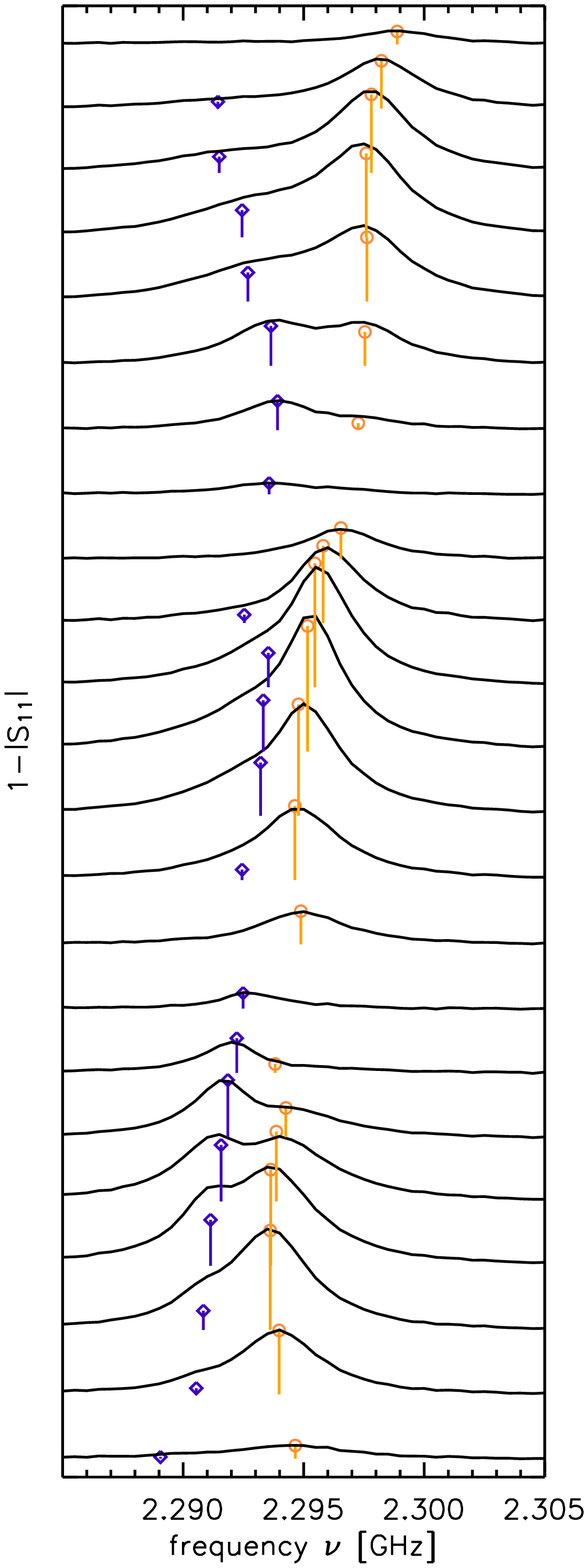}}
\end{minipage}\hspace{.1cm}
\begin{minipage}{50mm}\vspace*{-6ex}
 \parbox[t]{5mm}{(b)}\parbox[t]{50mm}{\ \\[-1ex] \includegraphics[width=50mm]{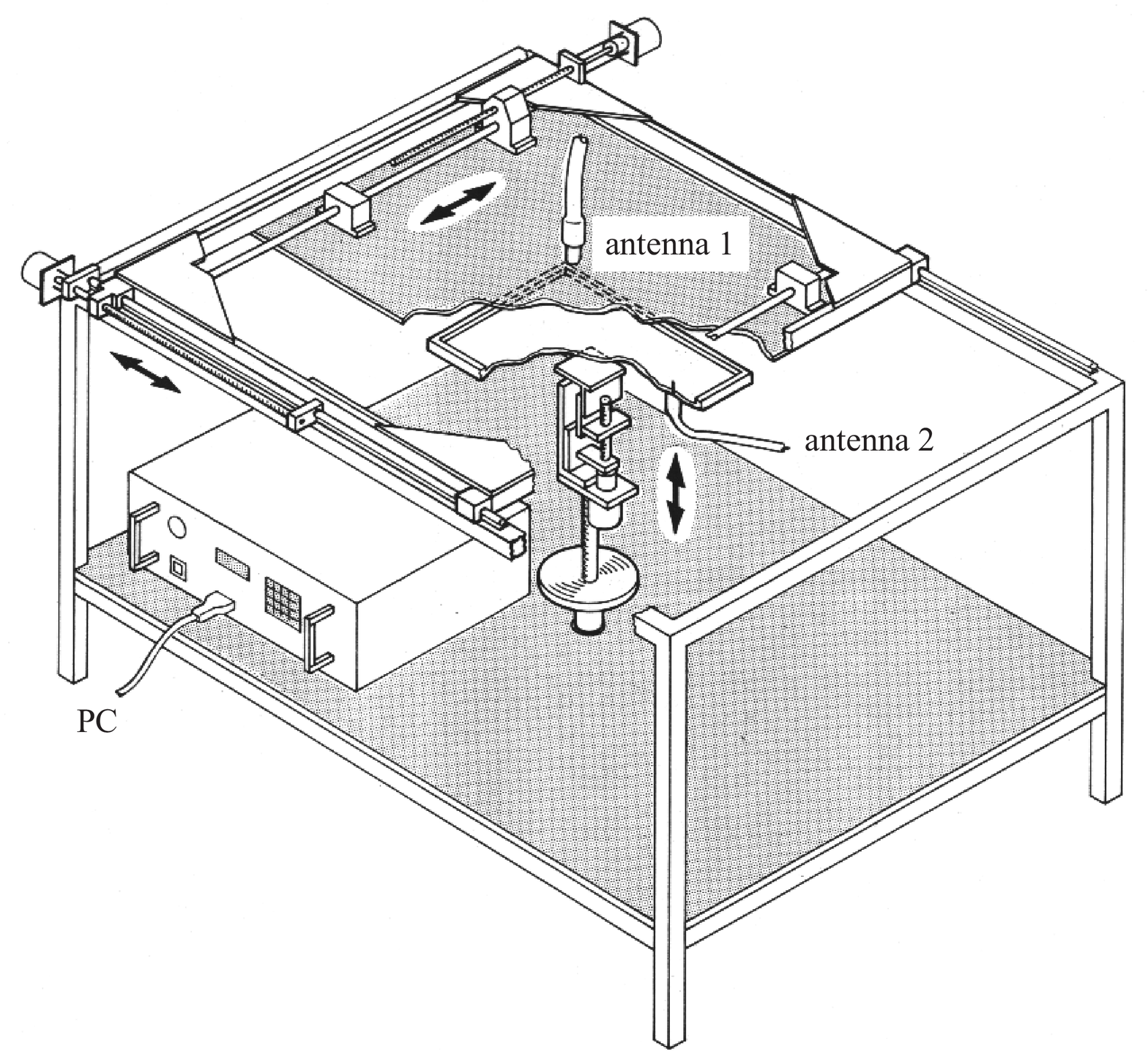}}
 \parbox[t]{5mm}{(c)}\parbox[t]{50mm}{\ \\[-1ex]\includegraphics[width=50mm]{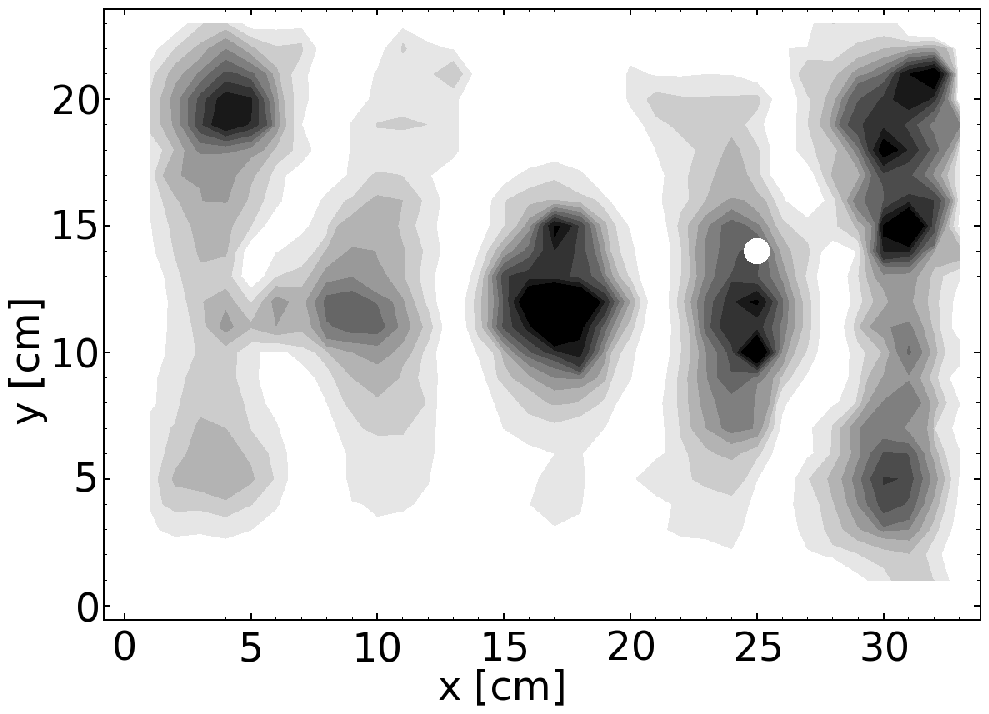}}
 \parbox[t]{5mm}{(d)}\parbox[t]{50mm}{\ \\[-1ex]\includegraphics[width=50mm]{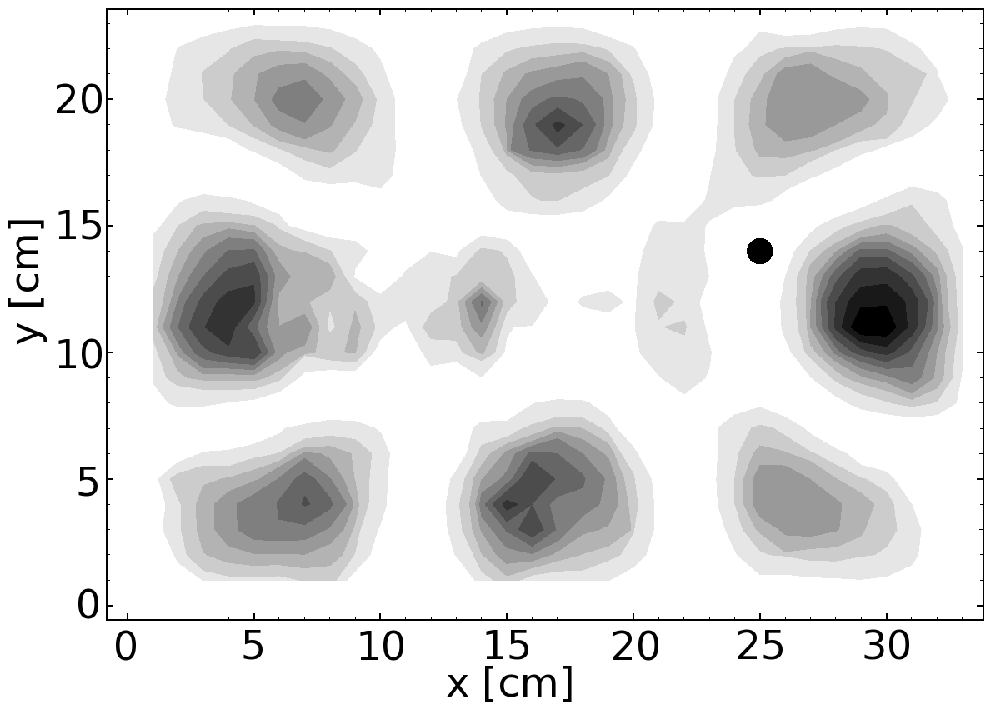}}
\end{minipage}
\end{center}
\caption{\label{fig:exp} Left (a): Experimental spectra of a rectangle for different positions of a movable antenna. The antenna is shifted by 10\,mm for each spectra in $y$ direction at a fixed $x$ = 5\,cm position. Diamond/circular marker gives the lower/high eigenfrequency. They are obtained by fitting a double Lorentzian. The lines below the correspond to the modulus of the height of the resonance. Right: At the top (b) a sketch of the measurement set-up is shown and below the two resulting eigenfunctions from the double Lorentz fit using the residua are plotted in grey scale. (c) for the lower and (d) for the higher frequency.}
\end{figure}

Figure~\ref{fig:exp}(b) shows a sketch of the apparatus allowing an automatic registration of microwave billiard wavefunctions. The billiard consists of two parts made of brass: \textit{i)} the upper part is a plate supporting a movable antenna in the center, \textit{ii)} the lower part is a cavity of a required shape (rectangular in our case), drilled-in into a thick plate. The apparatus was tested with help of a rectangular billiard with side lengths $a$ = 340\,mm, $b$ = 240\,mm, and a height of $h$ = 8\,mm. A \textit{fixed} antenna attached to the lower part at the position $x$ = 25\,cm, $y$ = 14\,cm (circular white/black markers in figure~\ref{fig:exp}(c,\,d)) allowed the registration of transmission spectra as well. Thus all components $S_{11}$, $S_{12}$, $S_{21}$, $S_{22}$ of the scattering matrix could be determined by the vector network analyzer. Each antenna consists of a small copper wire of diameter 0.3 mm attached to a standard SMA chassis connector and introduced through a small hole of diameter 1\,mm into the resonator.

Due to the presence of measuring antennas the resonances are shifted and broadened. As a consequence the non-overlapping resonance approximation becomes insufficient in the moment where shifts and widths induced by two (movable and fixed) antennas and additional broadenings due to the wall absorption are of the same order as the distance between the unperturbed resonances. In Figure~\ref{fig:exp}(a) the spectra for different $y$ antenna positions are shown in a frequency range where the perturbation by the antenna is of the order of the frequency distance between neighboring resonances. From the residua of each double Lorentzian fit the wave functions are obtained (see Figure~\ref{fig:exp}(c) and (d)). In these figures one sees that dark ``islands'' are connected with each other in several points via light ``bridges'', which evidently break the nodal lines of the pattern. The appearance of these bridges and the broken nodal lines of the patterns are the most important experimental imperfections in the case of overlapping resonances. We discuss these imperfections in details in Section \ref{sec:NearlyDegenerate}. In the billiard geometry described above we found several nearly degenerate states. The theoretical description used up to date \cite{kuh00b} had not been fully satisfactory, though describing some of the features of the obtained wavefunctions.
It is the purpose of this paper to obtain a full understanding of the experimental pattern, and especially the nodal line structures. The results are also valid if the perturbing bead method is used to measure wave functions \cite{sri91,lau94b,so95,doer98b,bog06}.

\section{\label{sec:sketch}Sketch of the general theory}

In this Section we remind of the general formulas we obtained in Part I of the paper. We have found the following expression for $S$-matrix of the cavity with attached antennas:
\begin{eqnarray}
S=-\frac{A-ik}{A+ik}+\frac{2ikB}{A+ik}\left[\hat G-\frac{BC}{A+ik}\right]^{-1}\frac{C}{A+ik},
\label{Smatrix1}
\end{eqnarray}
where $A,\,B,\,C$ are diagonal matrices with elements $A_i,\,B_i,\,C_i$ on diagonals and index $i$ denotes a number of an antenna. Here $C_i^*=B_i$ describe the coupling to outside and $A_i$ describes the boundary conditions at the end of the $i$-th antenna. The matrix $\hat G$ has elements
\begin{eqnarray}
\hat G_{ij}=\hat G_{ij}(\mathbf{R}_1,\ldots,\mathbf{R}_\nu;k^2)=\left\{\begin{array}{ll}
\xi_{\beta_i}(\mathbf{R}_i;k^2), & i=j, \\
G(\mathbf{R}_i,\mathbf{R}_j;k^2), & i\neq j.
\end{array}\right.
\end{eqnarray}
Here $\textbf{R}_i$, $i=1,\ldots\nu$ are the points of antennas attachments, $\xi_{\beta_i}(\mathbf{R}_i;k^2)$ is the renormalized Green function
\begin{equation}
\xi_{\beta_i}(\mathbf{R};k^2)=\lim_{\mathbf{r}\to\mathbf{R}}\left[G(\mathbf{r},\mathbf{R};k^2)+
\frac{1}{2\pi}\ln\left(\frac{|\mathbf{r}-\mathbf{R}|}{\beta_i}\right)\right]
\label{eq::xidef}
\end{equation}
and $\beta_i$ is the scattering length of the $i$-th antenna. Removing a global phase we can write the $S$-matrix in the form
\begin{equation}\label{eq::sm2}
S=1-\frac{2ikB}{A-ik}\left[\hat G-\frac{BC}{A+ik}\right]^{-1}\frac{C}{A+ik}.
\end{equation}
The resonances of the $S$-matrix \eref{eq::sm2} are obtained from the complex zeros of the equation
\begin{equation}
\textrm{det}\left(\hat G-BC/(A+ik)\right)=0.
\label{eq::detgen}
\end{equation}

\section{Wavefunction probing in case of overlapping resonances}
\label{sec:WaveProbing}

The advantage of the theory in Section~\ref{sec:sketch} is its universality as compared to the \textit{n}-poles approximation \cite{tud08}. Indeed, the expression for the $S$-matrix (\ref{eq::sm2}) is equally valid for closed cavities in the cases of isolated and overlapping resonances as well as for open cavities.

The $S$-matrix of a microwave cavity contains all available information \cite{stoe99}. Therefore any quantity of interest should be derived from the measured $S$-matrix. Being interested in wavefunction measurements we relate below the scattering matrix to the experimental wavefunction. However the experimental wavefunction in its turn can be easily related to the eigenmode of the closed cavity only in the case of an isolated resonance. It was already shown \cite{tud08} that for degenerate eigenvalues of the closed cavity one measures a sum of squares of all eigenfunctions corresponding to this eigenvalue. In this Section we go further and obtain a theoretical treatment of the experimental wavefunctions in the case of overlapping resonances.

To probe a wavefunction one uses a movable probing antenna and measures the reflection coefficient. If there are together $\nu$ antennas attached to a cavity, $\nu-1$ fixed and one movable, the reflection coefficient at the movable antenna is given by $S_{\nu\nu}$. Here the index $\nu$ is associated with the movable antenna. In the present case there are just two antennas, the fixed one in the bottom plate, and the moveable one in the upper part. From (\ref{eq::sm2}) we obtain
\begin{eqnarray}\label{eq::Sprobe}
  S_{\nu\nu}
  =1-2ik\frac{|B_\nu|^2}{A^2_\nu+k^2}\left[\hat G-\frac{BC}{A+ik}\right]_{\nu\nu}^{-1}.
\end{eqnarray}
Using the equalities
\begin{eqnarray}
\det[Q]=\det[K]\left(Q_{\nu\nu}-\sum_{i,j=1}^{\nu-1}Q_{\nu i}Q_{j \nu}K^{-1}_{ij}\right),\nonumber\\
Q^{-1}_{\nu\nu}=\left(Q_{\nu\nu}-\sum_{i,j=1}^{\nu-1}Q_{\nu i}Q_{j \nu}K^{-1}_{ij}\right)^{-1},
\end{eqnarray}
where $Q$ is an arbitrary invertible $\nu\times\nu$-matrix, and $K_{ij}=Q_{ij}$, $i,j=1,\ldots,\nu-1$, $K$ is invertible $(\nu-1)\times(\nu-1)$ matrix, we can rewrite the matrix element $[\hat G-BC/(A+ik)]_{\nu\nu}^{-1}$ in the following form
\begin{equation}
\fl
  \left[\hat G-\frac{BC}{A+ik}\right]_{\nu\nu}^{-1}=\frac{1}{\xi^{(\nu-1)}_{\beta_\nu}(\textbf{R}_\nu;k^2)-|B_\nu|^2/(A_\nu+ik)},
\label{eq::s11}
\end{equation}
where
\begin{equation}
\xi^{(\nu-1)}_{\beta_\nu}(\mathbf{R};k^2)=\lim_{\mathbf{r}\to\mathbf{R}}\left[G^{(\nu-1)}(\mathbf{r},\mathbf{R};k^2)
+\frac{1}{2\pi}\ln\left(\frac{|\mathbf{r}-\mathbf{R}|}{\beta_\nu}\right)\right].
\end{equation}
is the renormalized Green function of the system with $\nu-1$ fixed antennas \cite{tud08},
\begin{equation}
\fl
  G^{(\nu-1)}(\textbf{r},\textbf{R}_\nu;k^2)=G(\textbf{r},\textbf{R}_\nu;k^2)-\sum_{i,j=1}^{\nu-1}G(\mathbf{R}_\nu,\mathbf{R}_i;k^2)
  K^{-1}_{ij}G(\mathbf{R}_j,\mathbf{R}_\nu;k^2)
\end{equation}
is the Green function of the system with $\nu-1$ fixed antennas and $K_{ij}=[\hat G+D-BC/(A+ik)]_{ij}$, $i,j=1,\ldots,\nu-1$.

\subsection{Single pole approximation}
\label{subsec:SinglePol}

It was shown in \cite{tud08} that in the single pole approximation
\begin{equation}\label{eq::1pole}
  \left[\hat G-\frac{BC}{A+ik}\right]_{\nu\nu}^{-1}=V_\nu^{-1}-
  \frac{\psi^2_\mu(\textbf{R}_\nu)/V_\nu^2}{E_\mu-k^2+\sum_{i=1}^\nu \psi^2_\mu(\textbf{R}_i)/V_i},
\end{equation}
where $\psi_\mu(\textbf{r})$ is a normalized eigenfunction of a closed cavity, $E_\mu$ is the corresponding eigenvalue, and
\begin{equation}
  V_j=\lim_{k^2\to E_\mu}\left[\xi_{\beta_j}(\textbf{R}_j;k^2)-\frac{\psi_\mu^2(\textbf{R}_j)}{E_\mu-k^2}-\frac{|B_j|^2}{A_j+ik}\right]
\end{equation}
Equality (\ref{eq::1pole}) allows one to associate the residue of $S_{\nu\nu}$ with a modulo square of an eigenfunction of the closed cavity. The wavefunction can also be associated with a shift of resonance in the complex plane influenced by the motion of the $\nu$-th antenna
\begin{equation}
  k_\mu^2-E_\mu=\sum_{i=1}^{\nu-1} \psi^2_\mu(\textbf{R}_i)/V_i+\psi^2_\mu(\textbf{R}_\nu)/V_\nu.
\end{equation}
The described relations between $S_{\nu\nu}$ and wavefunctions of the closed cavity lead to two experimental standard techniques to measure wavefunctions. The wavefunction is either related to a residue or to a shift of a resonance. However both definitions are based on a single pole approximation and thus possess intrinsic imperfections. Below we analyze both methods in the case of overlapping resonances.

\subsection{Resonance shift}

Poles $k=k_\mu$ of (\ref{eq::s11}) can be found from the equation
\begin{equation}
  \xi^{(\nu-1)}_{\beta_\nu}(\textbf{R}_\nu;k)-|B_\nu|^2/(A_\nu+ik)=0.
\label{eq::nu-1}
\end{equation}
Equation (\ref{eq::nu-1}) can be easily treated if the pole structure of $\xi^{(\nu-1)}_{\beta_\nu}(\textbf{R}_\nu;k^2)$ is known. Let us find the residues of the function $\xi^{(\nu-1)}_{\beta_\nu}(\mathbf{R};k^2)$. To this end we will derive the convenient representation for $G^{(\nu-1)}(\mathbf{r},\mathbf{R};k^2)$. This Green function can be written in the form
\begin{eqnarray}
\fl
  G^{(\nu-1)}(\mathbf{r},\mathbf{R};k^2)&=&G^{(\nu-2)}(\mathbf{r},\mathbf{R};k^2)\nonumber\\
\fl
  &-&\frac{G^{(\nu-2)}(\mathbf{r},\mathbf{R}_{\nu-1};k^2)G^{(\nu-2)}(\mathbf{R}_{\nu-1},\mathbf{R};k^2)}
  {\xi^{(\nu-2)}_{\beta_{\nu-1}}(\mathbf{R}_{\nu-1};k^2)-|B_{\nu-1}|^2/(A_{\nu-1}+ik)},
\end{eqnarray}
where
\begin{eqnarray}
\fl
G^{(\nu-2)}(\mathbf{r},\mathbf{R};k^2)=G(\mathbf{r},\mathbf{R};k^2)-\sum_{i,j=1}^{\nu-2}G(\mathbf{r},\mathbf{R}_i;k^2)
  M^{-1}_{ij}G(\mathbf{R}_j,\mathbf{R};k^2)
\label{eq::G-2}
\end{eqnarray}
is the Green function of the system with $\nu-2$ antennas, $M_{ij}=Q_{ij}$, $i,j=1,\ldots,\nu-2$ (see \ref{app}). Thus the renormalized Green function $\xi^{(\nu-1)}(\mathbf{R}_\nu;k^2)$ takes the form
\begin{eqnarray}
  \fl
  \xi^{(\nu-1)}(\mathbf{R}_\nu;k^2)&=&\xi^{(\nu-2)}(\mathbf{R}_\nu;k^2)\nonumber\\
\fl
  &-&\frac{G^{(\nu-2)}(\mathbf{R}_\nu,\mathbf{R}_{\nu-1};k^2)G^{(\nu-2)}(\mathbf{R}_{\nu-1},\mathbf{R}_\nu;k^2)}
  {\xi^{(\nu-2)}_{\beta_{\nu-1}}(\mathbf{R}_{\nu-1};k^2)-|B_{\nu-1}|^2/(A_{\nu-1}+ik)}.
\label{eq::xi-1}
\end{eqnarray}
Now the $\mu$-th residue of $\xi^{(\nu-1)}(\mathbf{R}_\nu;k^2)$ can be found straightforwardly:
\begin{eqnarray}
\fl
  \textrm{Res}\,[\xi^{(\nu-1)}_{\beta_\nu}(\mathbf{R}_\nu;k^2)]_{k=q_\mu}\nonumber\\
\fl
  =\frac{G^{(\nu-2)}(\mathbf{R}_\nu,\mathbf{R}_{\nu-1};q_\mu^2)G^{(\nu-2)}(\mathbf{R}_{\nu-1},\mathbf{R}_\nu;q_\mu^2)}
  {[\partial\xi^{(\nu-2)}_{\beta_{\nu-1}}(\mathbf{R}_{\nu-1};k^2)/\partial k^2]_{k=q_\mu}+i|B_{\nu-1}|^2/2q_\mu(A_{\nu-1}+iq_\mu)^2},
\label{eq::resxi}
\end{eqnarray}
where $q_\mu$ is the $\mu$-th zero of the denominator in (\ref{eq::xi-1}).

It can be shown that $G^{(\nu-2)}(\mathbf{r},\mathbf{R}_{\nu-1};q_\mu^2)$ is an eigenfunction of the non-Hermitian operator corresponding to the cavity with $\nu-1$ attached antennas with boundary condition permitting only outgoing waves. In its turn $\bar G^{(\nu-2)}(\mathbf{R}_{\nu-1},\mathbf{r};q_\mu^2)=G^{(\nu-2)}(\mathbf{r},\mathbf{R}_{\nu-1};\bar q_\mu^2)$ is an eigenfunction of the corresponding Hermitian conjugate operator.

\subsection{Lorentz-fit}
\label{subsec:LorentzFit}

A Lorentz-fit of $S_{\nu\nu}$ obtained from the probing antenna gives residues corresponding to the poles $k^2_\mu$ of (\ref{eq::s11}). One associates the residue
\begin{eqnarray}\label{eq::res}
\fl
  \textrm{Res}\,\left[\hat G-\frac{BC}{A+ik}\right]_{\nu\nu}^{-1}=
  \frac{1}{[\partial\xi^{(\nu-1)}_{\beta_\nu}(\textbf{R}_\nu;k^2)/\partial k^2]_{k=k_\mu}+i|B_\nu|^2/2k_\mu(A_\nu+ik_\mu)^2},
\end{eqnarray}
where $k_\mu$ are solutions of (\ref{eq::nu-1}), with the modulo square of the $\mu$-th experimental eigenfunction.

\subsection{Small width approximation}
\label{subsec:SmallWidthApprox}

When $B_i\to 0$, $i=1,\ldots,\nu$ we can put $B_i=0$ in (\ref{eq::nu-1})-(\ref{eq::res}) but not in (\ref{eq::Sprobe}), since otherwise the resonance term in $S_{\nu\nu}$ would disappear. Then all poles becomes real and the $S$-matrix elements become singular. This approximation breaks immediately the unitarity of the $S$-matrix for a closed cavity. Nevertheless, it can be used since it almost preserves the residues as well as the positions of the resonances.

Then (\ref{eq::resxi}), (\ref{eq::res}) can be simplified. 
Indeed the Green function of a closed billiard with $\nu-1$ scatterers can be written in the form
\begin{equation}
  G^{(\nu-1)}(\textbf{r},\textbf{R};k^2)=\sum_{\mu=1}^\infty
  \frac{G^{(\nu-2)}(\textbf{r},\textbf{R}_{\nu-1};q_\mu^2)G^{(\nu-2)}(\textbf{R}_{\nu-1},\textbf{R};q_\mu^2)}
  {N_\mu^2(q_\mu^2-k^2)}.
\end{equation}
where
\begin{equation}\label{eq::nmu}
  N_\mu^2=-\frac{\partial\xi^{(\nu-2)}_{\beta_{\nu-1}}}{\partial k^2}(\textbf{R}_{\nu-1};q_\mu^2).
\end{equation}
Using the equality following from (\ref{eq::xidef}) and the standard spectral sum we find
\begin{equation}
\fl
  \frac{\partial\xi_\beta^{(j)}}{\partial k^2}(\textbf{R};k^2)=\lim_{\mathbf{r}\to\mathbf{R}}
  \frac{\partial G^{(j)}(\textbf{r},\textbf{R};k^2)}{\partial k^2}=-\int_{\sigma} d^2r \left[G^{(j)}(\textbf{r},\textbf{R};k^2)\right]^2,
\end{equation}
where $\sigma$ is the area of the billiard. Thus
\begin{eqnarray}
   \frac{\partial \xi^{(\nu-2)}_{\beta_{\nu-1}}}{\partial k^2}=-\int_{\sigma} d^2r \left[G^{(\nu-2)}(\textbf{r},\textbf{R};k^2)\right]^2\label{eq::xi-nu-2},\\
   \frac{\partial \xi^{(\nu-1)}_{\beta_\nu}}{\partial k^2}=-\int_{\sigma} d^2r \left[G^{(\nu-1)}(\textbf{r},\textbf{R};k^2)\right]^2\label{eq::xi-nu-1}.
\end{eqnarray}

From \eref{eq::resxi} and \eref{eq::xi-nu-2} we conclude that the renormalized Green function $\xi^{(\nu-1)}$ takes the form \cite{tud1}
\begin{eqnarray}
\fl
  \xi^{(\nu-1)}_{\beta_\nu}(\textbf{R};k^2)&=&\xi^{(\nu-1)}_{\beta_\nu}(\textbf{R};0)
  +\sum_{\mu=1}^\infty f_\mu^2(\textbf{R})\left(\frac{1}{q_\mu^2-k^2}-\frac{1}{q_\mu^2}\right),
\label{eq::zerow}
\end{eqnarray}
where $k=q_\mu$ are the solutions of the equation
\begin{equation}
  \xi^{(\nu-2)}_{\beta_{\nu-1}}(\textbf{R}_{\nu-1};k^2)=0
\end{equation}
and
\begin{equation}
   f_\mu(\textbf{r})=G^{(\nu-2)}(\textbf{r},\textbf{R}_{\nu-1};q_\mu^2)/N_\mu
\end{equation}
are normalized eigenfunctions of the system with $\nu-2$ scatterers.

From \eref{eq::res} and \eref{eq::xi-nu-1} we find that the pattern associated with a residue is proportional to
\begin{equation}
  \frac{1}{\int_{\sigma} d^2r \left[G^{(\nu-1)}(\textbf{r},\textbf{R};k_\mu^2)\right]^2},
\end{equation}
where $k=k_\mu$ are the solutions of the equation
\begin{equation}
  \xi^{(\nu-1)}_{\beta_\nu}(\textbf{R}_\nu;k^2)=0.
\end{equation}

\section{Experimental patterns in the case of nearly degenerate states}
\label{sec:NearlyDegenerate}

The universal form of the representation (\ref{eq::zerow}) allows one to analyze main features of experimental wave-patterns equally for any number of introduced antennas. Especially simple analysis can be performed if the experimental patterns are associated with changes of resonance positions while the probing antenna is moving inside a cavity. We perform this analysis for the case of nearly degenerate states.

In \cite{tud08} we have shown that in the case of a degenerate state the observed pattern is a sum of squares of all wavefunctions forming the orthogonal basis in the subspace corresponding to the degenerate eigenvalue. Thus an experimentally measured pattern has no longer a simple relation to an individual wavefunction of the unperturbed system.

What is happening in the situation with nearly degenerate states? It is clear that when a shift of a resonance due to a presence of a measuring antenna is comparable with the level-spacings, one can not treat an observable wave-pattern as a square of an isolated wavefunction. One can expect at least that the nodal lines of both eigenfunctions remain in the experimental patterns, since an eigenstate can not be excited by an antenna moving along its nodal line. However this expectation fails. We show below that in this situation the measured wavefunctions are very peculiar: they cannot only be related to isolated wavefunctions, but the nodal lines of these patterns are broken.

For our purpose it is sufficient to consider the renormalized Green function (\ref{eq::zerow}) in two poles approximation. Substituting it into (\ref{eq::nu-1}) with $B_\nu=0$ we obtain
\begin{equation}
\frac{f_1^2(\mathbf{R})}{z_1-k^2}+\frac{f_2^2(\mathbf{R})}{z_2-k^2}
+D(\textbf{R})=0,
\label{2psAppr}
\end{equation}
where $\mathbf{R}=\textbf{R}_\nu$ is the position of the probing antenna, $z_\mu=q_\mu^2$,
\begin{eqnarray}
  f_i(\textbf{R})=\frac{G^{(\nu-2)}(\textbf{R},\textbf{R}_{\nu-1};z_\mu)}{N_\mu},\\
  D(\textbf{R})=\left[\xi^{(\nu-1)}_{\beta_\nu}(\textbf{R};k^2)
  -\frac{f_1^2(\mathbf{R})}{z_1-k^2}-\frac{f_2^2(\mathbf{R})}{z_2-k^2}\right]_{k^2=(z_1+z_2)/2}.
\end{eqnarray}
We assume that $|D(\textbf{R})|$ is large enough to neglect the influence of all other poles. Under this assumption (\ref{2psAppr}) has two positive solutions $k^2=k_1^2,\,k_2^2$.

Consider first the possibility of the degeneracy of two solutions $k_1^2=z_1,\,k_2^2=z_2$ of \eref{2psAppr}. The condition $k_1^2(\mathbf{R})=k_2^2(\mathbf{R})$ seems to define a curve in the plane of the cavity. However from the form of \eref{2psAppr} one can see that there are always two different solutions except for $f_1(\mathbf{R})=0$ and $k^2(\textbf{R})=z_1$ (analogously if $f_2(\mathbf{R})=0$ and $k^2(\textbf{R})=z_2$). Thus the degeneracy can occur only if a scatterer is placed somewhere at a nodal line.

Since the degeneracy of eigenvalues can occur only at isolated points, in the vicinity of these points eigenvalues should split. Consider the nodal line $\mathbf{r}=\mathbf{R}(\tau)$ of the function $f_1(\mathbf{r})$:
$f_1(\mathbf{R}(\tau))=0$. Along this line \eref{2psAppr} reads
$f_2(\mathbf{R}(\tau))/[z_2-k^2(\tau)]+D[\mathbf{R}(\tau)]=0.$
Let us assume that $k^2(\tau)=z_1$ for some value of the parameter $\tau=\tau_0$. At the point $\mathbf{R}_0=\mathbf{R}(\tau_0)$ the eigenvalue $z_1$ is twofold degenerate.

\begin{figure}
\begin{center}
\begin{minipage}{120mm}
 \parbox{7mm}{(a)}\parbox{35mm}{\includegraphics[width=35mm]{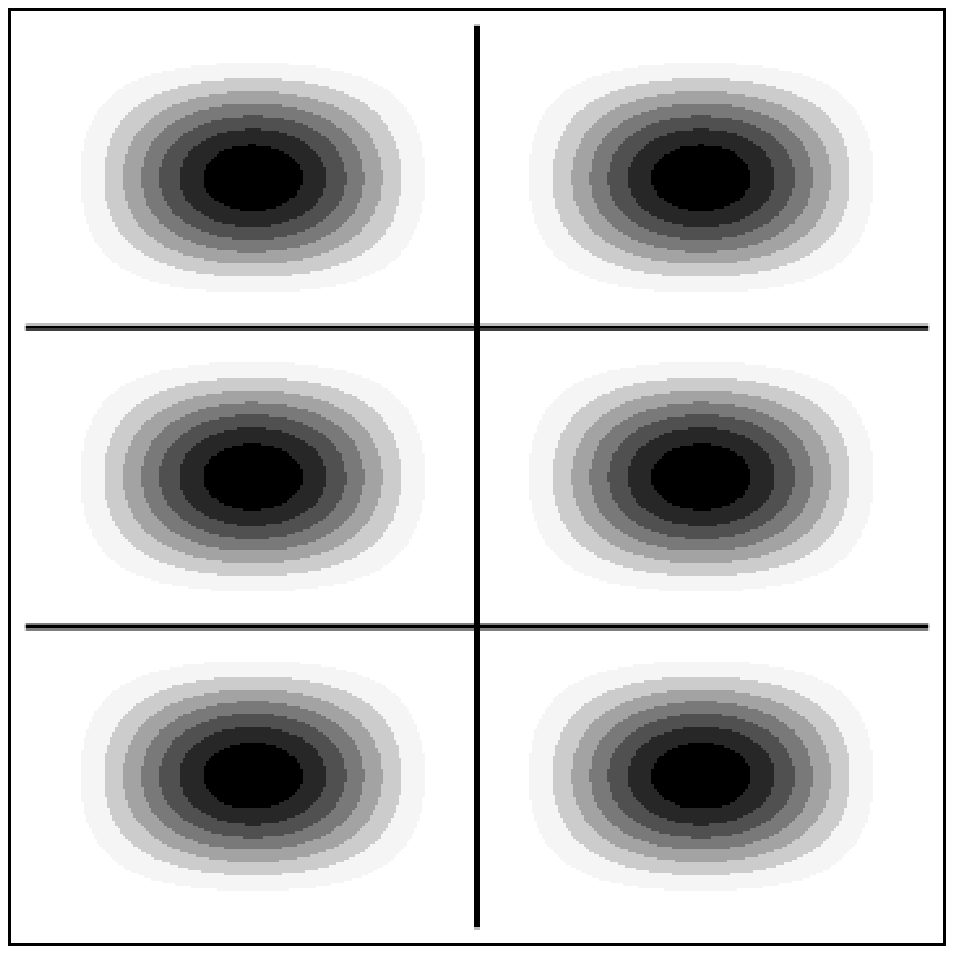}}
 \parbox{35mm}{\includegraphics[width=35mm]{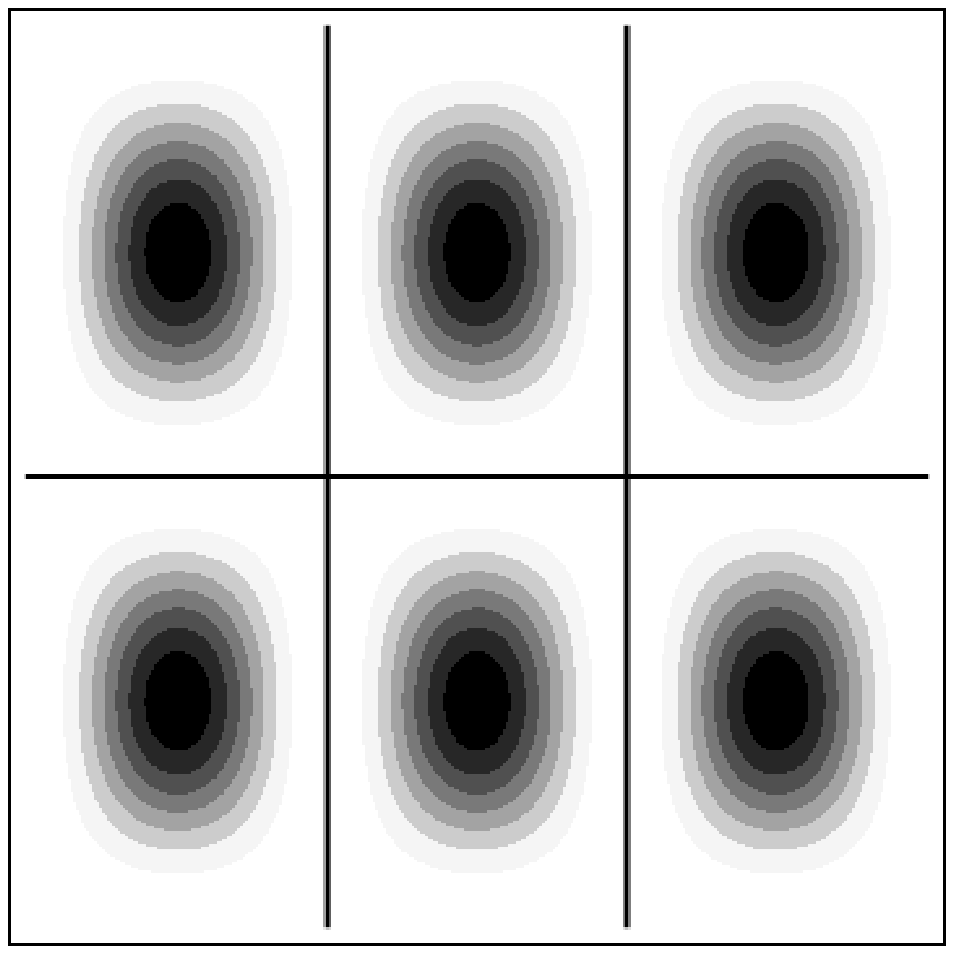}}\\[.3cm]
 \parbox{7mm}{(b)}\parbox{35mm}{\includegraphics[width=35mm]{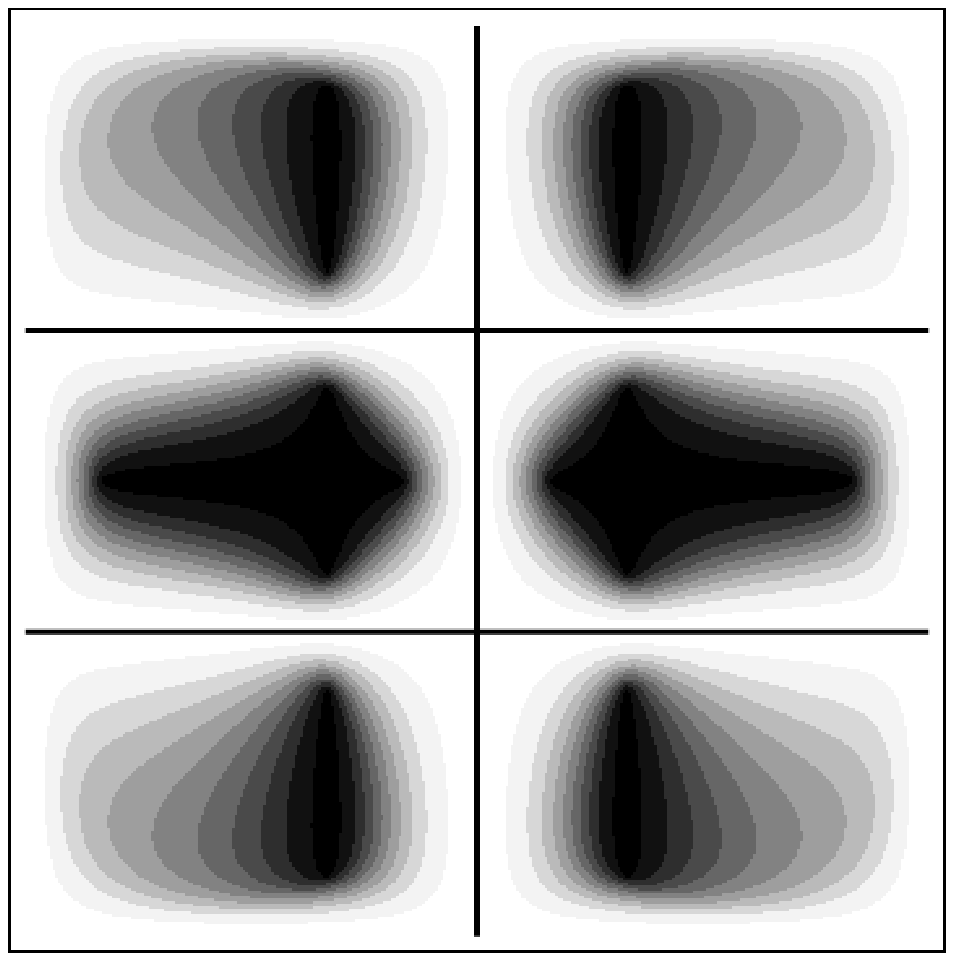}}
 \parbox{35mm}{\includegraphics[width=35mm]{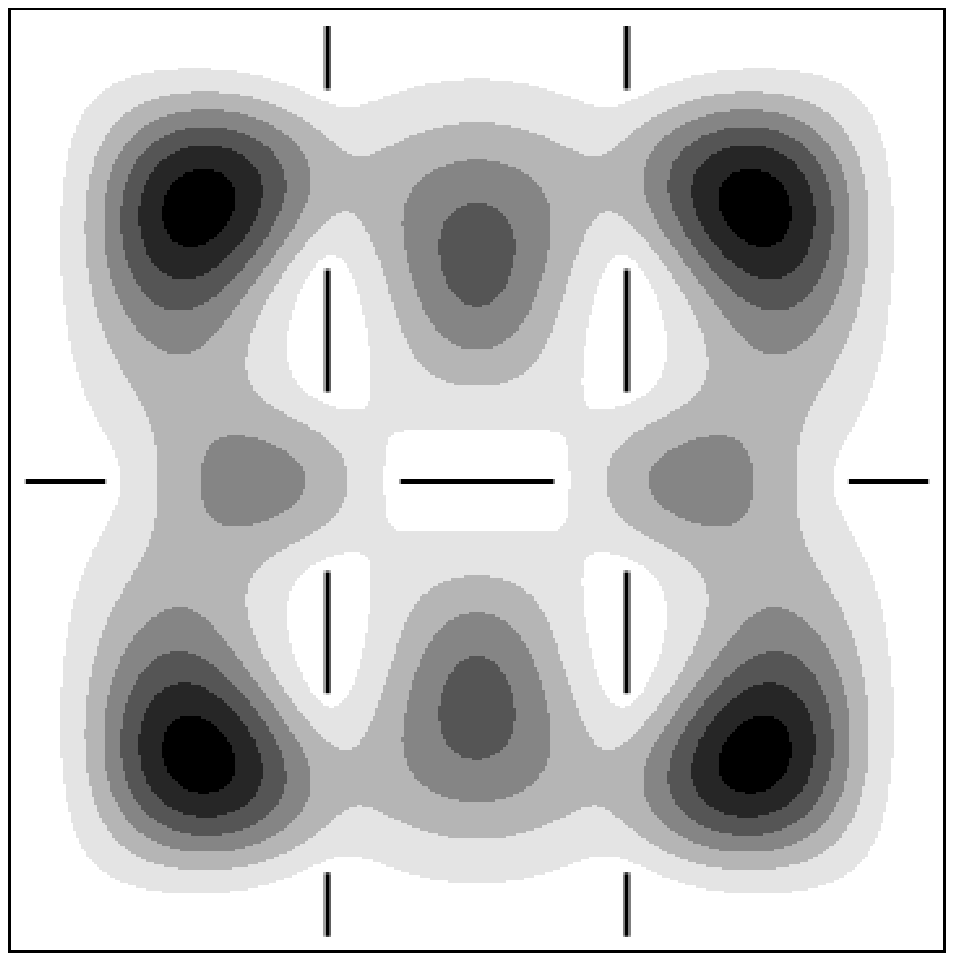}}
 \parbox{35mm}{\rotatebox{270}{\includegraphics[width=35mm]{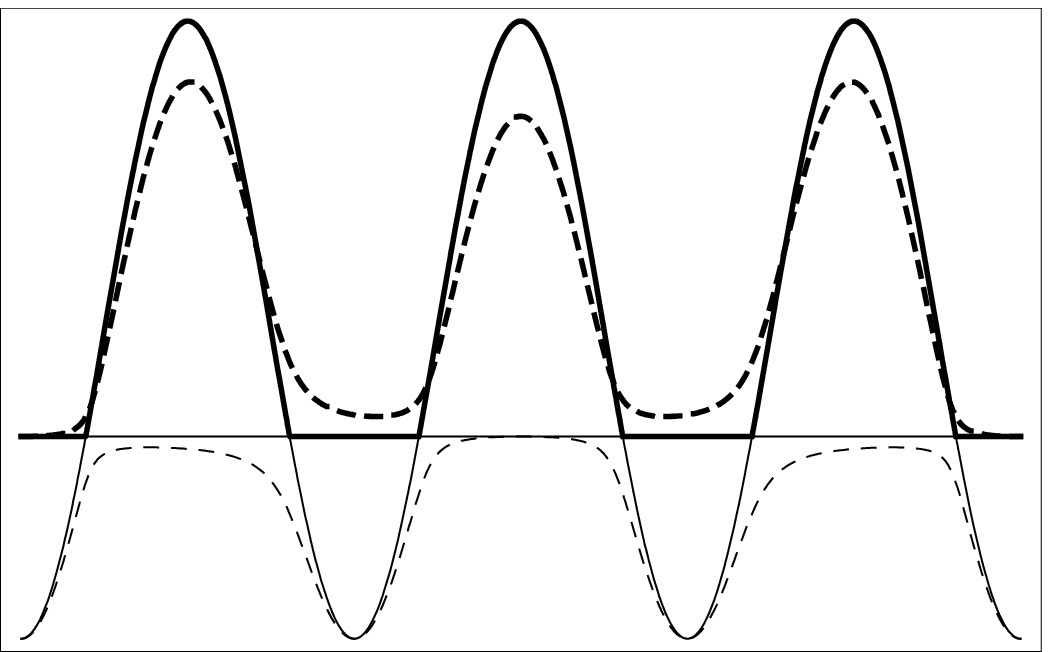}}}\\[.3cm]
 \parbox{7mm}{(c)}\parbox{35mm}{\includegraphics[width=35mm]{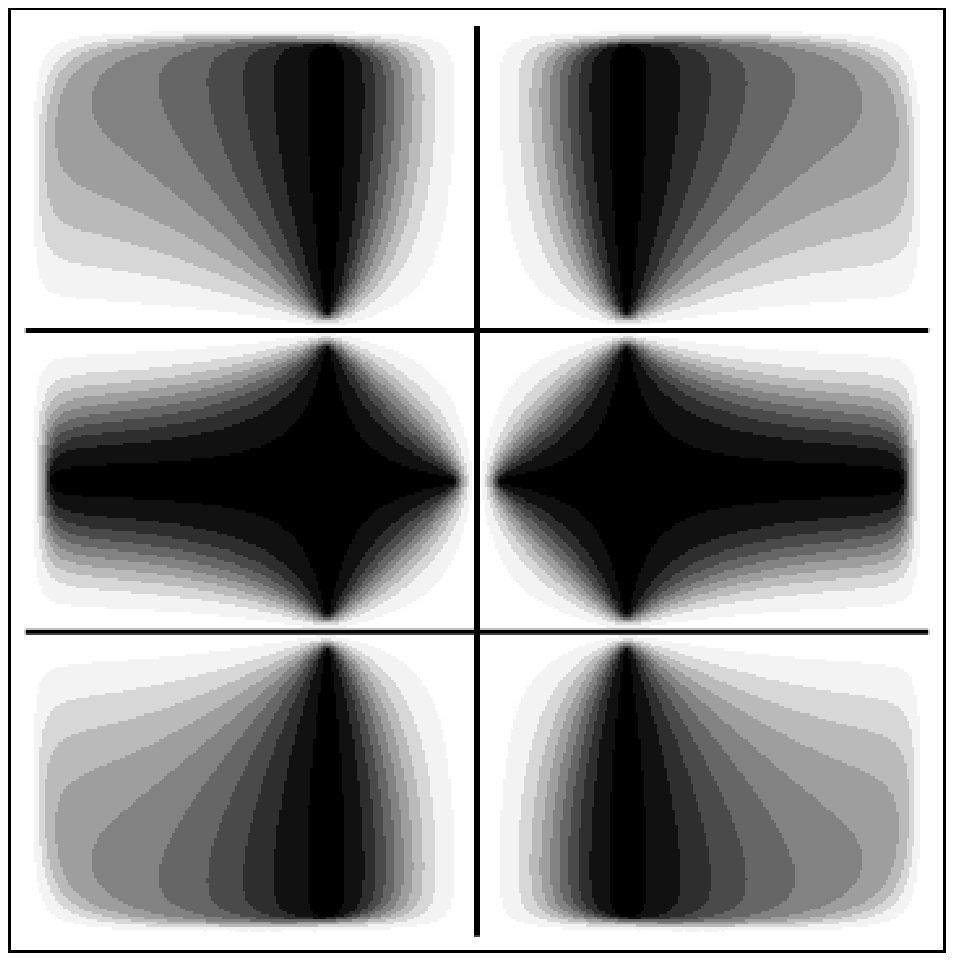}}
 \parbox{35mm}{\includegraphics[width=35mm]{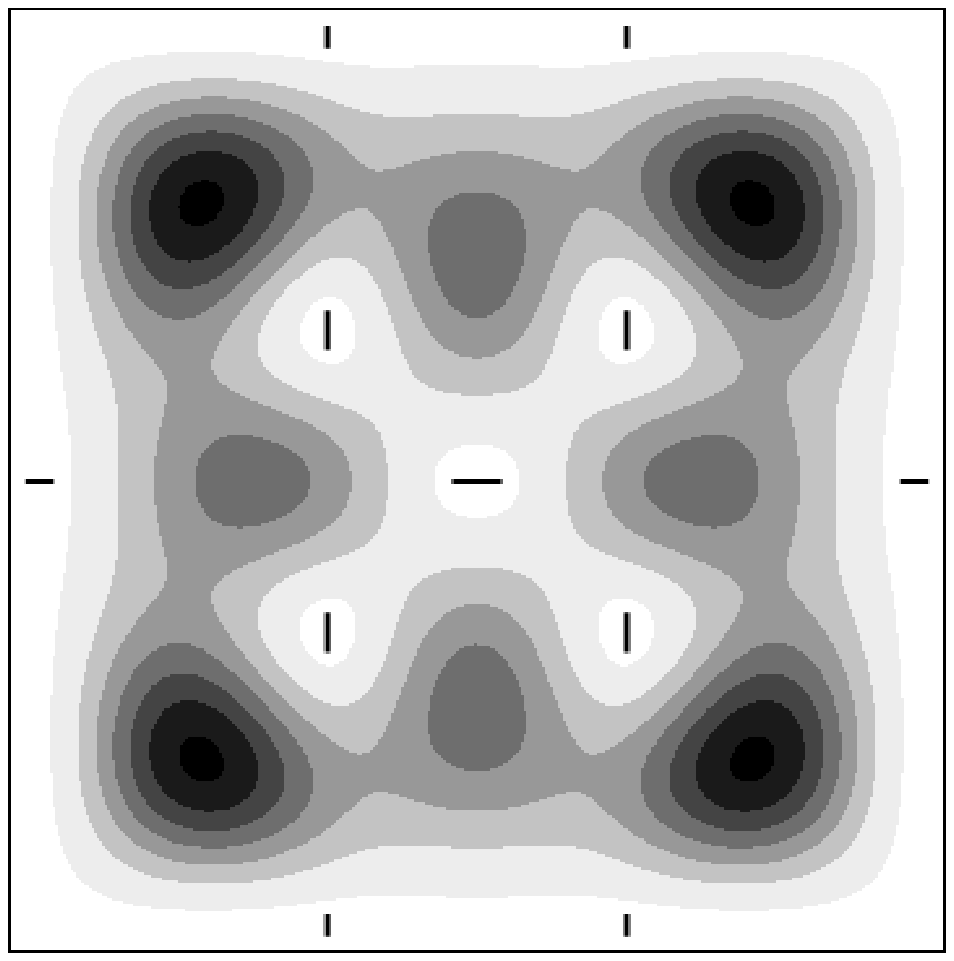}}
 \parbox{35mm}{\rotatebox{270}{\includegraphics[width=35mm]{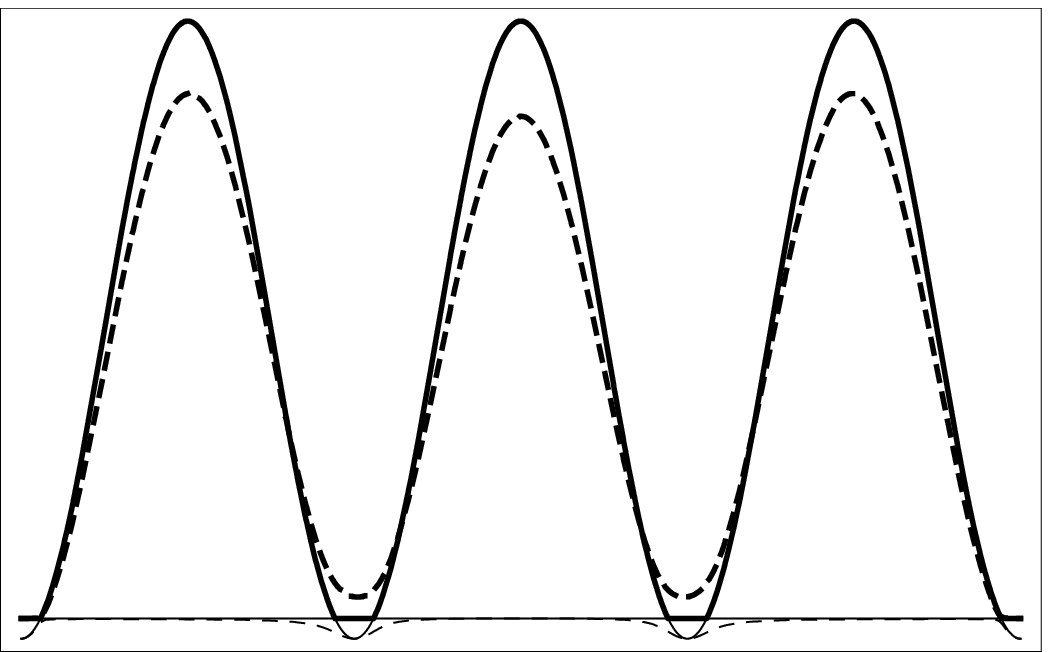}}}\\[.3cm]
 \parbox{7mm}{(d)}\parbox{35mm}{\ } \parbox{35mm}{\includegraphics[width=35mm]{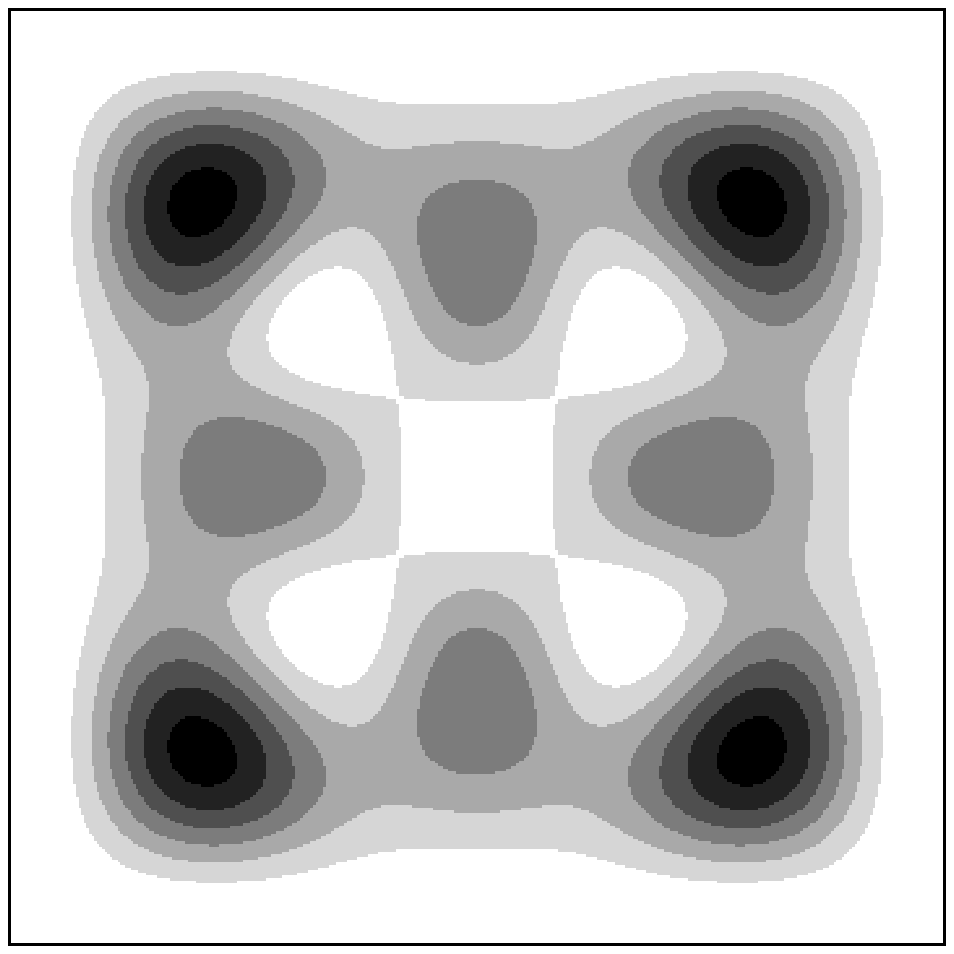}}
\end{minipage}
\end{center}
\caption{\label{fig::pat1}
The expected experimental patterns (squares of experimental wavefunctions) in the case of two nearly degenerate states of a rectangular billiard. The side ratio $d_x/d_y$ of the rectangle is $1/1.01$. Row (a) shows a  pair of eigenfunctions of the rectangle with quantum numbers (2,3) and (3,2), respectively. The next two rows show the evolution of the experimental patterns corresponding to the higher and to the lower eigenvalue, respectively, for two different position independent perturbations D=1 (b) and 0.l (c). The nodal lines are inserted into the pattern as black solid lines for better visuality. In the left column all nodal lines remain unperturbed, in the right one they become interrupted (b,c) and tend to points for (d). The right column show shifts of two resonances (\textit{x}-axes) along the left vertical nodal line in Figures (b, c) (solid curves) and a line close and parallel to it (interrupted curves). In figure (d) the sum of figures (a) is plotted.}
\end{figure}

Let us now consider the line $\textbf{r}=\textbf{R}(\tau)+\boldsymbol{\xi}(\tau)$, where $|\boldsymbol{\xi}(\tau)|>0$ is small enough. The solutions $k^2_{1,2}$ of (\ref{2psAppr}) along this line read
\begin{eqnarray}
\fl
k^2_{1,2}=\frac{ D(\mathbf{R}+\boldsymbol{\xi})(z_1+z_2)+f_1^2(\mathbf{R}+\boldsymbol{\xi})+
f_2^2(\mathbf{R}+\boldsymbol{\xi})\pm \sqrt{\mathcal{D}(\mathbf{R}+\boldsymbol{\xi})}}{2 D(\mathbf{R}+\boldsymbol{\xi})},\\
\fl
\mathcal{D}(\textbf{R})=\Bigl[D(\mathbf{R})(z_1-z_2)+f_1^2(\mathbf{R})-f_2^2(\mathbf{R})
\Bigr]^2
+4f_1^2(\mathbf{R})f_2^2(\mathbf{R})
\label{eq::k212}
\end{eqnarray}
where plus and minus signs correspond to different branches, since $\mathcal{D}>0$. Far from the point $\textbf{R}=\textbf{R}_0$ we can write
\begin{equation}
k^2_{1,2}\simeq\frac{ D(\mathbf{R})(z_1+z_2)+f_2^2(\mathbf{R})\pm
| D(\mathbf{R})(z_1-z_2)-f_2^2(\mathbf{R})|}{2 D(\mathbf{R})}
\end{equation}
or
\begin{eqnarray}
\fl
  k^2_{1}\simeq z_1, &\quad k^2_2=z_2+\frac{f_2^2(\mathbf{R})}{ D(\mathbf{R})}, &\quad \textrm{if} \quad \frac{f_2^2(\mathbf{R})}{z_2-z_1}+ D(\mathbf{R})>0,\nonumber\\
\fl
  k^2_{1}\simeq z_2+\frac{f_2^2(\mathbf{R})}{ D(\mathbf{R})}, &\quad k^2_2=z_1, &\quad \textrm{if} \quad \frac{f_2^2(\mathbf{R})}{z_2-z_1}+ D(\mathbf{R})<0.
\end{eqnarray}
In the vicinity of the point $\textbf{R}=\textbf{R}_0$ the expression
\begin{equation}
\frac{f_2^2(\mathbf{R})}{z_2-z_1}+ D(\mathbf{R})
\end{equation}
changes its sign. Thus in the vicinity of this point along the line $\textbf{r}=\textbf{R}(\tau)+\boldsymbol{\xi}(\tau)$ an avoided crossing takes place.

Figure~\ref{fig::pat1} illustrates what may happen in a wavefunction measurement in the case of nearly degenerated eigenfunctions. The calculation has been performed for a rectangle with a side ratio of 1/1.01, i.\,e. very close to the square. Figures (a) show the eigenfunction with quantum numbers $(2,3)$ and $(3,2)$, respectively, for the rectangle without perturbation by an antenna. Due to the small side mismatch there is an energy splitting of the the eigenstates of the rectangle.

The avoided crossing allows us to separate properly the patterns corresponding to lower and higher frequencies (see figure~\ref{fig::pat1}\, (c)). It also shows that a nodal line of an unperturbed eigenfunction may become interrupted (figure \ref{fig::pat1}\,(b,c) center column), and the ``bridges'' connecting nodal domains through the nodal line may appear. These bridges are expected to appear in the case when a shift of a resonance due to an antenna or a perturber is larger than the level-spacing.

On the contrary, the structure of nodal lines in the first column in figure \ref{fig::pat1} remains. This is clear since for both resonances the shifts due to the presence of a scatterer has always the same sign. When the influence of the perturbation increases ($| D(\mathbf{R})|$ decreases), the pattern keeping the nodal line structure of the original unperturbed state is concentrating in vicinities of nodal lines of the the other unperturbed state. The other pattern tends to the sum of squares of two unperturbed wavefunctions (see figure \ref{fig::pat1}\,(d)), i.e. tends to the pattern expected for a degenerate state\cite{tud08}.

One can see that measured nodal lines for the case of a single scatterer are continuous if the coupling of a scatterer is weak as compared to level spacing. For a stronger coupling nodal lines become interrupted and turn into the points of intersection of nodal lines of $f_1(\mathbf{r})$ and $f_2(\mathbf{r})$.

\subsection{Experimental verification}

\begin{figure}
\begin{center}
\begin{minipage}{50mm}
 \parbox{5mm}{(a)}\parbox{50mm}{\includegraphics[width=50mm]{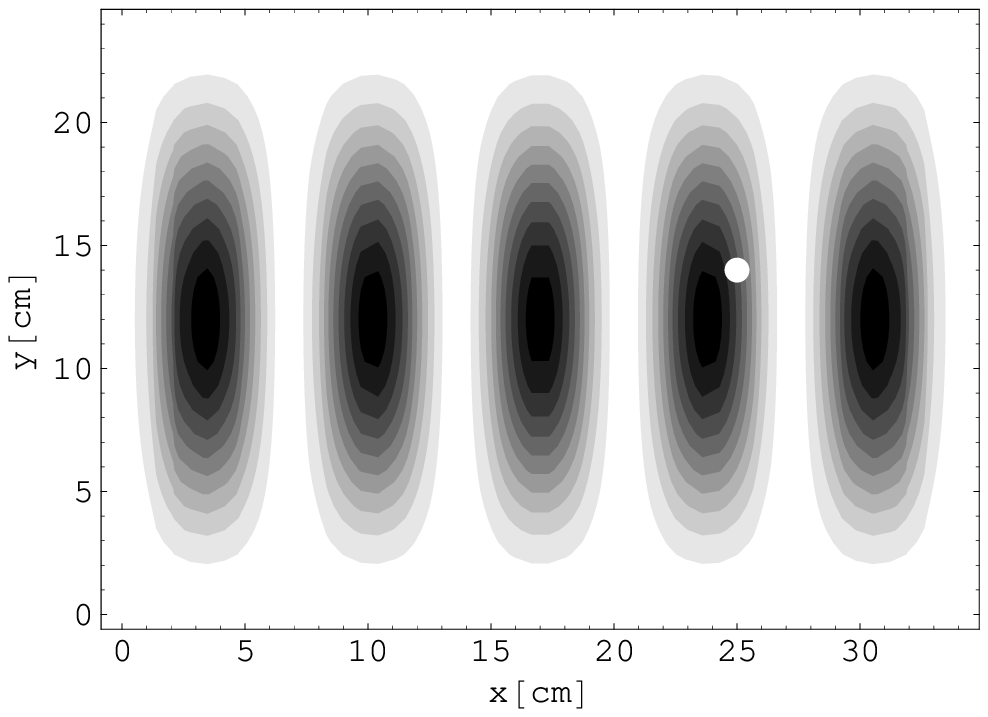}}
 \parbox{5mm}{(c)}\parbox{50mm}{\includegraphics[width=50mm]{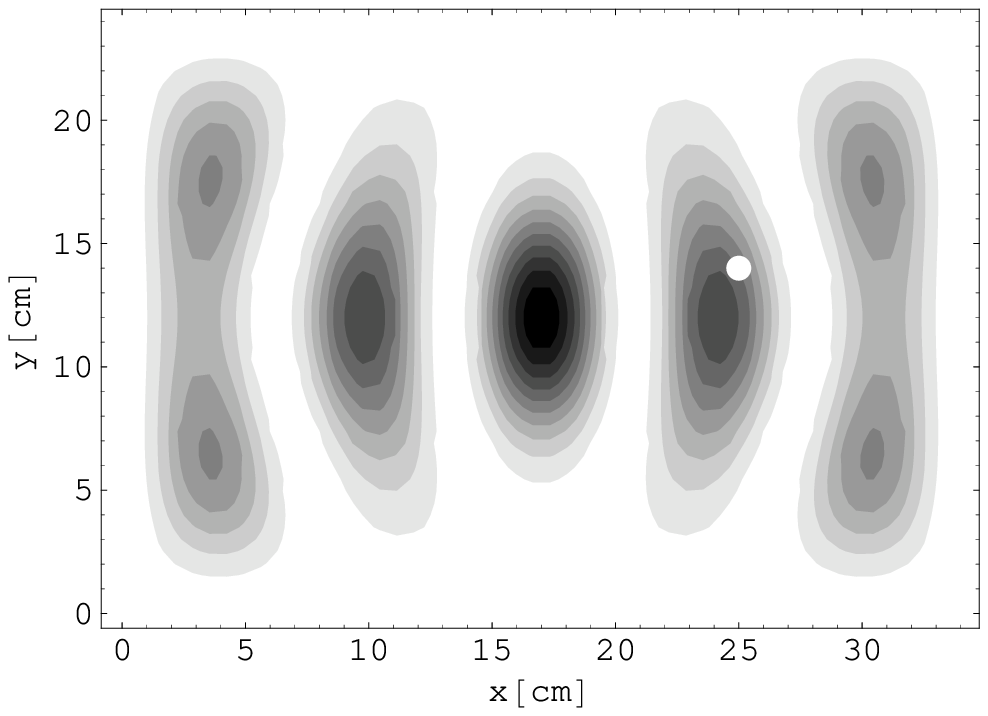}}
 \parbox{5mm}{(e)}\parbox{50mm}{\includegraphics[width=50mm]{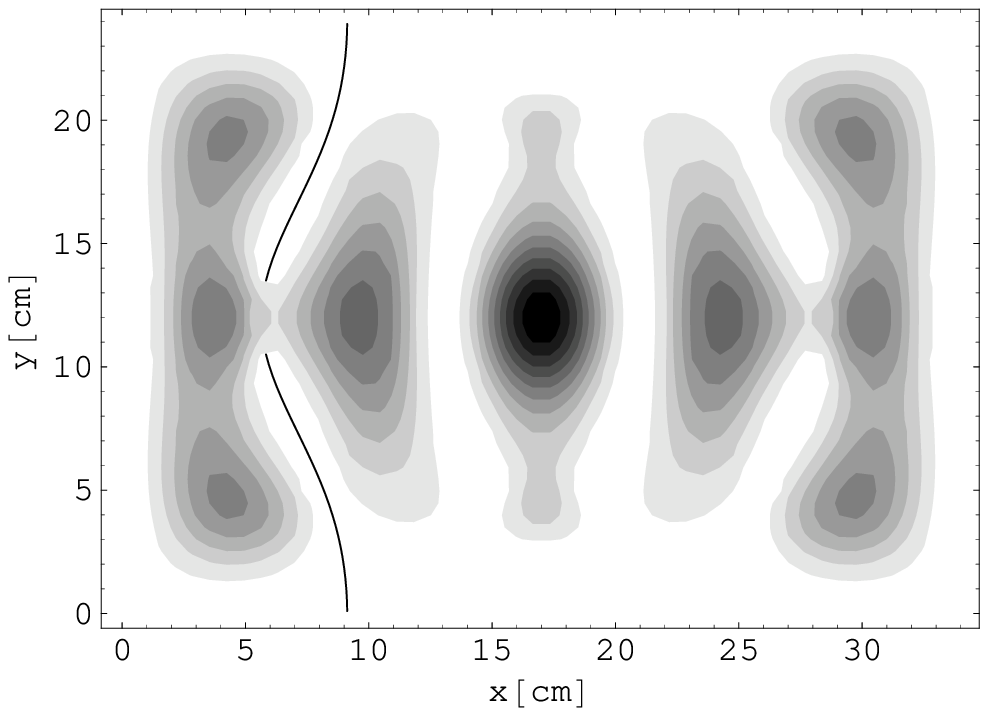}}
 \parbox{5mm}{(g)}\parbox{50mm}{\includegraphics[width=50mm]{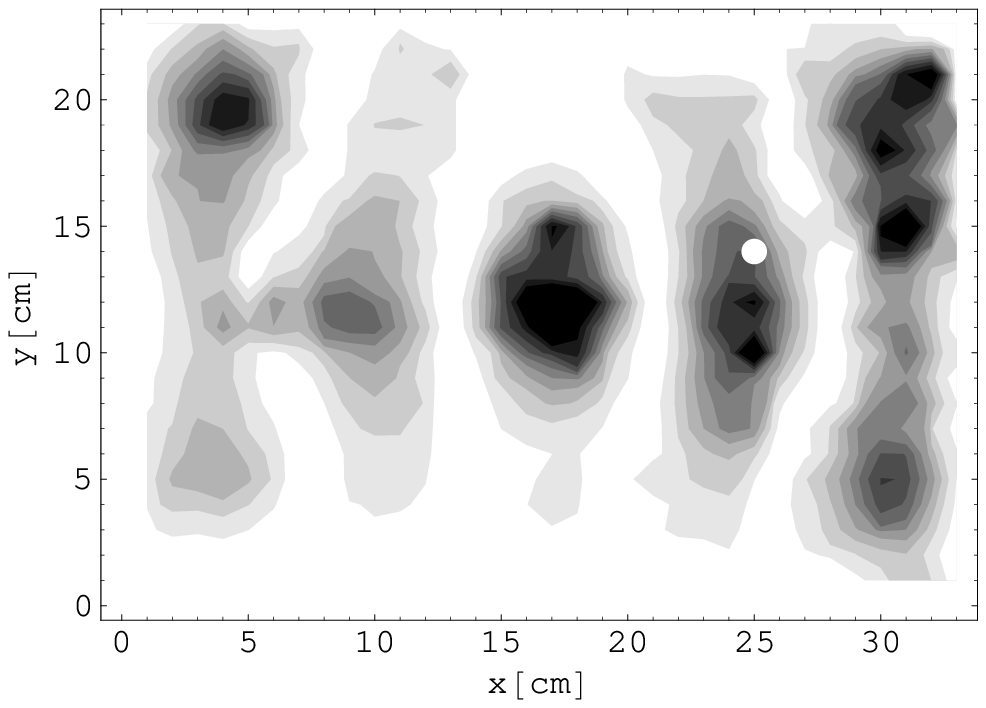}}
\end{minipage}\hspace{1cm}
\begin{minipage}{50mm}
 \parbox{5mm}{(b)}\parbox{50mm}{\includegraphics[width=50mm]{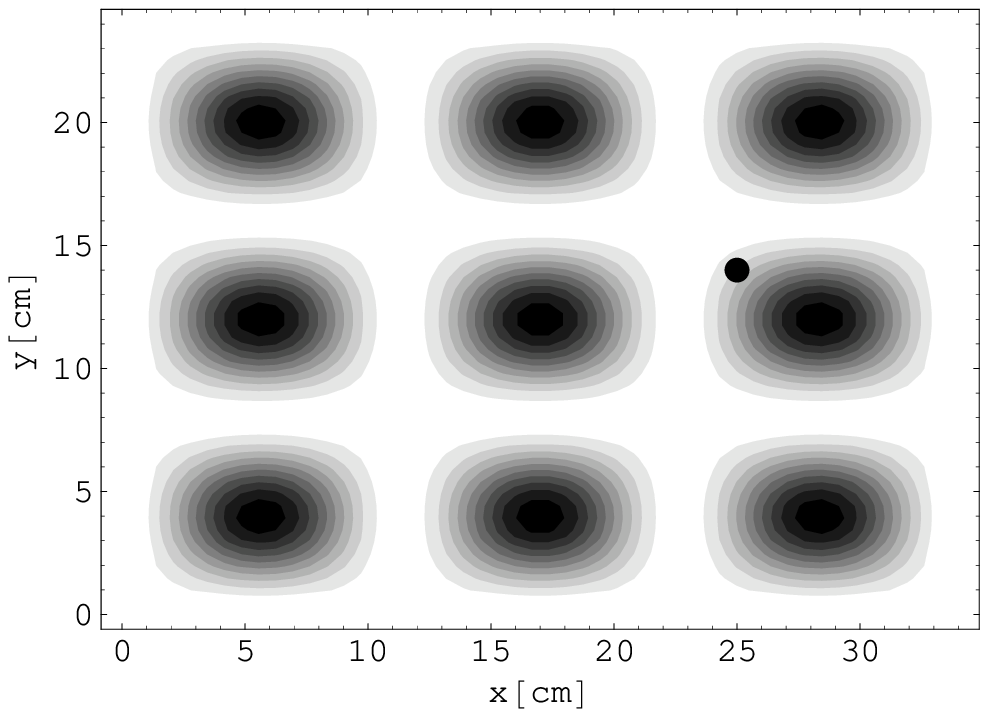}}
 \parbox{5mm}{(d)}\parbox{50mm}{\includegraphics[width=50mm]{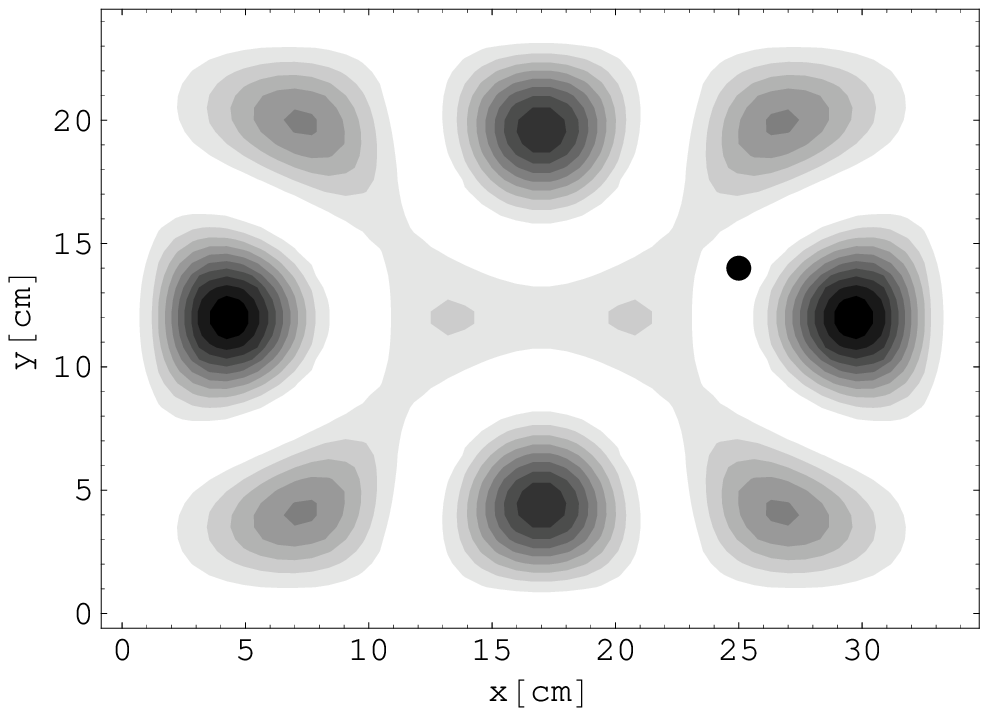}}
 \parbox{5mm}{(f)}\parbox{50mm}{\includegraphics[width=50mm]{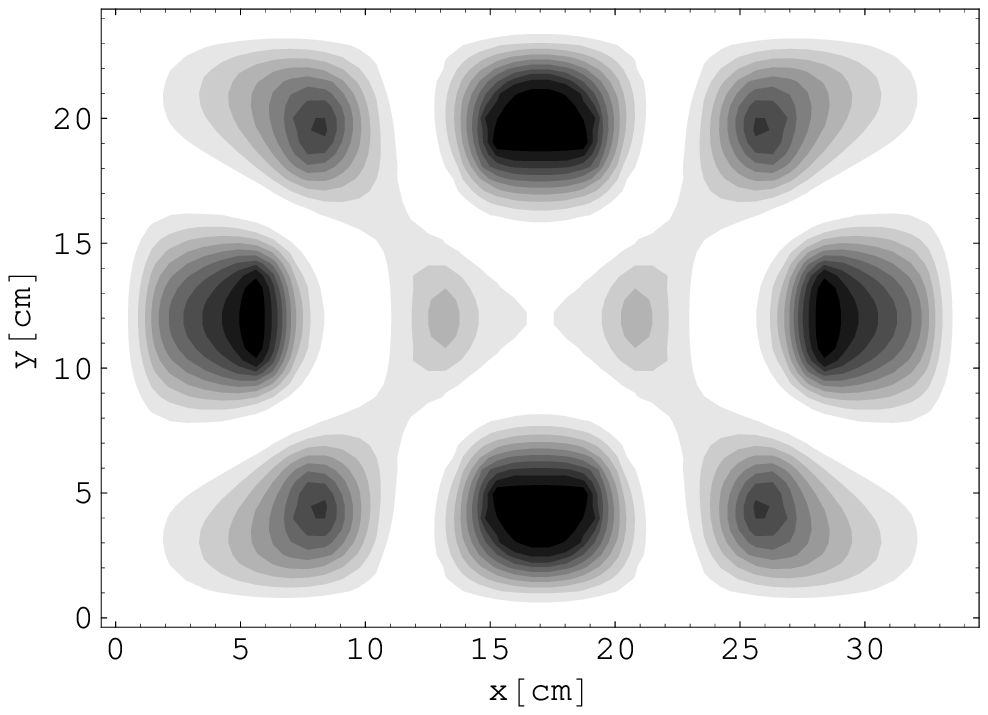}}
 \parbox{5mm}{(h)}\parbox{50mm}{\includegraphics[width=50mm]{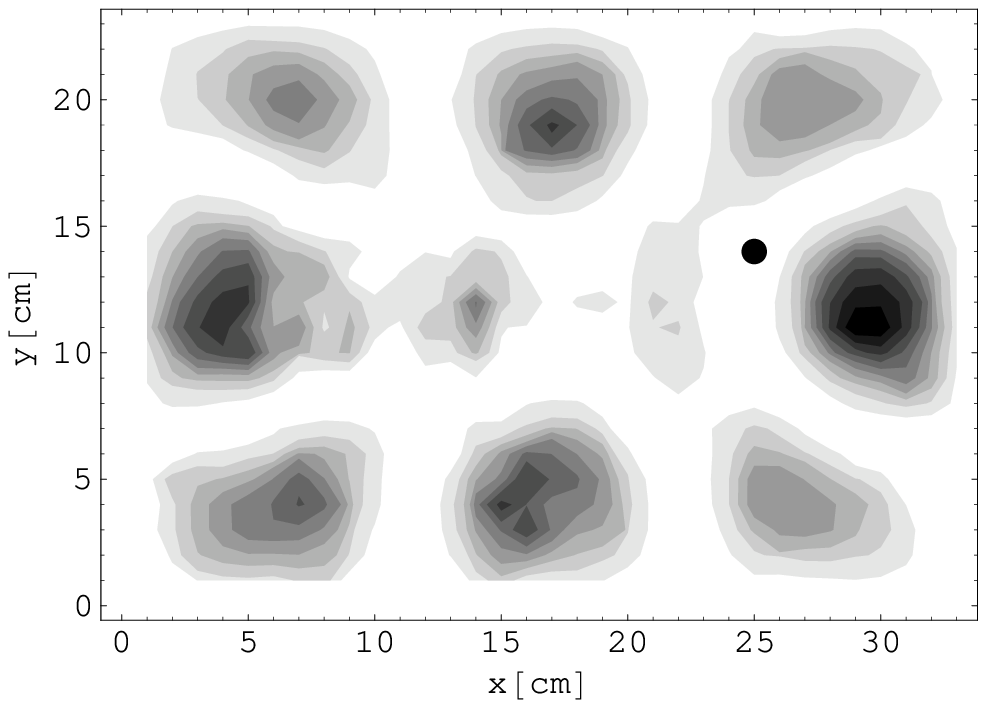}}
\end{minipage}
\end{center}
\caption{\label{efrb} Calculated distribution $|\psi_{nm}|^2$. White and black points show the position of the fixed antenna $x=25$cm, $y=14$cm. a,b) Without antennas: $|\psi_{nm}|^2$: a) $n=5,\,m=1,\,\omega_{51}=2.29113$GHz; b) $n=3,\,m=3,\,\omega_{33}=2.29348$GHz. Here and further $k[\textrm{cm}^{-1}]\simeq 0.2096\omega[\textrm{GHz}]$.
c) $|G(\mathbf{r},\mathbf{R},k^2)|^2$, $\omega=2.28706$GHz.
d) The combination of eigenfunctions a) and b) equal to zero at the antenna point.
e,f) Simulation of the experimental wavefunctions in the billiard with two coupled antennas using the Ewald's method~\cite{tud1} and the standard procedure of the Lorenzian fit \cite{ste92}. We used the following parameters: $A_1=A_2=0.3\,\textrm{dm}^{-1}$, $B_1=B_2=2.5\,\textrm{dm}^{-1/2}$, $D_1=-25$, $D_2=-65$.
g,h) Experimentally measured wavefunctions: g) $\omega=2.298$GHz, h) $\omega=2.307$GHz.}
\end{figure}

Now we have all the insights to treat the experimental findings presented in Section~\ref{Sec:Exp}. The experiment was performed on the rectangular billiard with sides $d_x=34$cm and $d_y=24$cm. The square of the side ratio $(d_y/d_x)^2=(34/24)^2\simeq 2$. The eigenvalues of the unperturbed billiard are
\begin{equation}
E_{nm}=\pi^2[(n/d_x)^2+(m/d_y)^2]\simeq \pi^2 d_x^2[n^2+2m^2].
\end{equation}
It is easy to show that $n^2+2m^2$ can be degenerate. An example is $n_1=5,\,m_1=1,\,n_2=3,\,m_2=3$.

Figure \ref{efrb} illustrates the comparison of numerical simulations with experimentally measured patterns corresponding to the eigenstates with quantum numbers $n=5,\,m=1$ and $n=3,\,m=3$. In numerical simulations we assumed for simplicity that  $D_1$, $ D_2$ do not depend on $\mathbf{R}$. We also assumed that the coupling strength of both antennas is negligible, thus the resonances can be found from zero-width approximation.

In the figures \ref{efrb}a), \ref{efrb}b) one sees the original eigenfunctions of the cavity. In figures \ref{efrb}c), \ref{efrb}d) we neglected the splitting of eigenvalues as well as the influence of the movable antenna. Figures \ref{efrb}e), \ref{efrb}f) illustrate the numerical simulation with fixed and movable antennas. Plotted values are the solutions $k^2$ of (\ref{eq::nu-1}). Finally figures \ref{efrb}g), \ref{efrb}h) represent the experimental data. One can see that there is a reasonable coincidence between experimental figures and the numerics.

This work was supported by the Deutsche Forschungsgemeinschaft via an individual grant and the Forschergruppe 760: Scattering systems with complex dynamics.

\appendix

\section{\label{app}Representations of the Krein's resolvent}

Krein's resolvent $G^{(\nu)}(\textbf{r},\textbf{R};k^2)$ of the billiard with $\nu$ antennas reads
\begin{equation}\label{eq::a1}
\fl
  G^{(\nu)}(\textbf{r},\textbf{R};k^2)=G(\textbf{r},\textbf{R};k^2)-
  \sum_{i,j=1}^{\nu}G(\mathbf{r},\mathbf{R}_i;k^2)
  Q^{-1}_{ij}G(\mathbf{R}_j,\mathbf{R};k^2).
\end{equation}
From the physical point of view it is obvious that instead of the unperturbed billiard corresponding to the free Green function $G(\textbf{r},\textbf{R};k^2)$ one can consider the billiard with $m<\nu$ antennas corresponding to the Green function $G^{(m)}(\textbf{r},\textbf{R};k^2)$ as the unperturbed one and the rest of $\nu-m$ antennas as the perturbation. Then (\ref{eq::a1}) can be rewritten in the form
\begin{equation}\label{eq::a2}
\fl
  G^{(\nu)}(\textbf{r},\textbf{R};k^2)=G^{(m)}(\textbf{r},\textbf{R};k^2)-
  \sum_{i,j=1}^{\nu-m}G^{(m)}(\mathbf{r},\mathbf{R}_i;k^2)
  \left[Q^{(m)}_{ij}\right]^{-1}G^{(m)}(\mathbf{R}_j,\mathbf{R};k^2),
\end{equation}
where $Q^{(m)}_{ij}=[\hat G^{(m)}+D-BC/(A+ik)]_{ij}$, $i,j=1,\ldots,\nu$,
\begin{eqnarray}
\hat G^{(m)}_{ij}=\hat G^{(m)}_{ij}(\mathbf{R}_1,\ldots,\mathbf{R}_\nu;k^2)=\left\{\begin{array}{ll}
\xi^{(m)}_{\beta_i}(\mathbf{R}_i;k^2), & i=j, \\
G^{(m)}(\mathbf{R}_i,\mathbf{R}_j;k^2), & i\neq j.
\end{array}\right.\\
\xi^{(m)}_{\beta_i}(\mathbf{R};k^2)=\lim_{\mathbf{r}\to\mathbf{R}}\left[G^{(m)}(\mathbf{r},\mathbf{R};k^2)+
\frac{1}{2\pi}\ln\left(\frac{|\mathbf{r}-\mathbf{R}|}{\beta_i}\right)\right].
\label{eq::ximdef}
\end{eqnarray}
The formal proof follows from the equality
\begin{eqnarray}
\fl
  \left(
  \begin{array}{cc}
    Q_{11} & Q_{12} \\
    Q_{21} & Q_{22} \\
  \end{array}\right)^{-1}\nonumber\\
\fl\qquad
    =\left(\begin{array}{cc}
    Q^{-1}_{11}+Q^{-1}_{11}Q_{12}\left[Q^{(m)}\right]^{-1}Q_{21}Q^{-1}_{11} & -Q^{-1}_{11}Q_{12}\left[Q^{(m)}\right]^{-1} \\
    -\left[Q^{(m)}\right]^{-1}Q_{21}Q_{11}^{-1} & \left[Q^{(m)}\right]^{-1} \\
  \end{array}\right),
\label{eq::a5}
\end{eqnarray}
valid for any invertible matrices $Q_{11},\ Q_{22}$. Here $Q_{11}$ is $m\times m$, $Q_{22}$ is $(\nu-m)\times(\nu-m)$ matrix, $Q^{(m)}=Q_{22}-Q_{21}Q_{11}^{-1}Q_{12}$.
Substituting (\ref{eq::a5}) into (\ref{eq::a1}) we find
\begin{eqnarray}
\fl
  G^{(\nu)}(\textbf{r},\textbf{R};k^2)=G(\textbf{r},\textbf{R};k^2)-
  \langle\textbf{r},m|Q^{-1}_{11}|\textbf{R},m\rangle\nonumber\\
\fl
  \qquad-\Bigl(\langle\textbf{r},\nu-m|-\langle\textbf{r},m|Q^{-1}_{11}Q_{12}\Bigr)\left[Q^{(m)}\right]^{-1}\Bigl(|\textbf{R},\nu-m\rangle-Q_{21}Q^{-1}_{11}|\textbf{R},m\rangle\Bigr),
\label{eq::a6}
\end{eqnarray}
where $\langle\textbf{r},\nu-m|_i=G(\textbf{r},\textbf{R}_i;k^2)$, $\langle\textbf{r},m|_j=G(\textbf{r},\textbf{R}_j;k^2)$,
$|\textbf{R},\nu-m\rangle_i=G(\textbf{R}_i,\textbf{R};k^2)$, $|\textbf{R},m\rangle_j=G(\textbf{R}_j,\textbf{R};k^2)$, $i=1,\ldots,\nu-m$, $j=1,\ldots,m$. Taking into account equalities
\begin{eqnarray}
\Bigl(\langle\textbf{r},\nu-m|-\langle\textbf{r},m|Q^{-1}_{11}Q_{12}\Bigr)_i=G^{(m)}(\textbf{r},\textbf{R}_i;k^2),\nonumber\\
\Bigl(|\textbf{R},\nu-m\rangle-Q_{21}Q^{-1}_{11}|\textbf{R},m\rangle\Bigr)_j=G^{(m)}(\textbf{R}_j,\textbf{R};k^2),
\end{eqnarray}
we conclude that (\ref{eq::a6}) is identical with (\ref{eq::a2}).

\vspace{1cm}


\end{document}